\documentclass[12pt]{article}

\usepackage{graphicx}
\usepackage{psfrag}

\def\beq{\begin{equation}}
\def\eeq{\end{equation}}
\def\bea{\begin{eqnarray}}
\def\eea{\end{eqnarray}}
\def\nn{\nonumber}
\def\bra#1{\left\langle #1\right|}
\def\ket#1{\left| #1\right\rangle}
\def\roughly#1{\mathrel{\raise.3ex\hbox
{$#1$\kern-.75em\lower1ex\hbox{$\sim$}}}}
\def\lsim{\roughly<}

\def\sss{\scriptscriptstyle}

\def\bd{B_d^0}
\def\bdbar{{\overline{B_d^0}}}
\def\bs{B_s^0}
\def\bsbar{{\overline{B_s^0}}}
\def\ks{K_{\sss S}}
\def\kbar{{\bar K}^0}
\def\btod{{\bar b} \to {\bar d}}
\def\btos{{\bar b} \to {\bar s}}
\def\ANPq{{\cal A}^q}
\def\ANPu{{\cal A}^u}
\def\ANPd{{\cal A}^d}
\def\pewc{P_{\sss EW}^{\sss C}}
\def\pewcp{P_{\sss EW}^{\prime \sss C}}
\def\plb#1#2#3{{ Phys.\ Lett.} {\bf #1B}, #3 (#2)}
\def\prd#1#2#3{{ Phys.\ Rev.} {\bf D#1}, #3 (#2)}

\def\prl#1#2#3{{ Phys.\ Rev.\ Lett.} {\bf #1}, #3 (#2)}

\newcommand{\av}[1]{\langle #1 \rangle}

\begin{document}

\begin{flushright}
UdeM-GPP-TH-04-126 \\
McGill 20/04 \\
UAB-FT-573 \\
\end{flushright}

\begin{center}
\bigskip
{\Large \bf \boldmath $\bd(t)\to \pi^+\pi^-$ and $\bs(t)\to K^+ K^-$ 
Decays: \vskip2truemm 
A Tool to Measure New-Physics Parameters} \\
\bigskip
\bigskip
{\large David London $^{a,b,}$\footnote{london@lps.umontreal.ca},
Joaquim Matias $^{c,}$\footnote{matias@ifae.es} and Javier Virto
$^{c,}$\footnote{jvirto@ifae.es}}
\end{center}

\begin{flushleft}
~~~~~~~~~~~$a$: {\it Laboratoire Ren\'e J.-A. L\'evesque,
Universit\'e de Montr\'eal,}\\
~~~~~~~~~~~~~~~{\it C.P. 6128, succ. centre-ville, Montr\'eal, QC,
Canada H3C 3J7} \\
~~~~~~~~~~~$b$: {\it Physics Department, McGill University,}\\
~~~~~~~~~~~~~~~{\it 3600 University St., Montr\'eal QC, Canada H3A 2T8}\\
~~~~~~~~~~~$c$: {\it IFAE, Universitat Aut\`onoma de Barcelona,}\\
~~~~~~~~~~~~~~~{\it 08193 Bellaterra, Barcelona, Spain}
\end{flushleft}

\begin{center}
\bigskip (\today)
\vskip0.5cm
{\Large Abstract\\}
\vskip3truemm
\parbox[t]{\textwidth} {If physics beyond the standard model is
present in $B$ decays, experimental measurements seem to suggest that
it principally affects those processes with significant $\btos$
penguin amplitudes. It was recently argued that, in general, such
new-physics (NP) effects can be parametrized in terms of a single NP
amplitude $\ANPq$ and phase $\Phi_q$, for $q=u,d,s,c$. In this paper,
we show that the study of the decays $\bs(t)\to K^+ K^-$ and
$\bd(t)\to \pi^+\pi^-$ allows one to measure the NP parameters $\ANPu$
and $\Phi_u$. We examine the implications for this method of the
latest experimental results on these decays. If NP is found in
$\bs(t)\to K^+ K^-$, it can be partially identified through
measurements of $\bs(t)\to K^0 \kbar$.}
\end{center}

\thispagestyle{empty}
\newpage
\setcounter{page}{1}
% Decrease texheight (for preprint numbers) again
%\textheight 23.0 true cm
\baselineskip=14pt

To date, theoretical analyses of CP violation in the $B$ system have
generally concentrated on two subjects. First, many different methods
have been proposed for extracting the CP-violating angles $\alpha$,
$\beta$ and $\gamma$ of the unitarity triangle \cite{pdg} within the
standard model (SM) \cite{CPreview}. Second, there have been numerous
studies of new-physics (NP) signals through measurements of CP
violation in $B$ decays. We now have many ways of detecting the
presence of NP. However, there has been relatively little work on the
third and final ingredient, which is to find ways of {\it measuring}
the NP parameters. If this can be done it might be possible to
identify the new physics, before the production of new particles at
high-energy colliders.

A first step in this direction was taken in Ref.~\cite{DLNP}, in which
it was shown that one can reduce the number of NP parameters to a
manageable level, and measure them. The knowledge of these parameters
then allows one to partially identify the new physics. The argument
goes as follows. 

At present, we have a number of experimental hints of new physics.
First, within the SM, the CP asymmetry in $\bd(t) \to J/\psi \ks$
($\sin 2\beta$) is equal to that measured in other decays dominated by
the quark-level transition $\btos s{\bar s}$. However, Belle finds a
discrepancy of $2.2\sigma$ between the CP asymmetry in $\bd(t) \to
\phi\ks$ and that in $\bd(t) \to J/\psi\ks$ \cite{Sakai}. In addition,
the BaBar measurement of the CP asymmetry in $\bd(t)\to \eta'\ks$
differs from that in $\bd(t) \to J/\psi \ks$ by $3.0\sigma$
\cite{Giorgi}. Second, ratios of various $B\to \pi K$ branching
ratios, which are equal in the SM \cite{KpiBRs}, are found to differ
from one another by $1.6\sigma$ \cite{HFAG}. Furthermore, Belle finds
a $2.4\sigma$ discrepancy with the SM in $B\to \pi K$ direct
asymmetries: $A_{\sss CP} (K^+ \pi^-) \ne A_{\sss CP} (K^\pm \pi^0)$,
a result confirmed by BaBar \cite{Sakai}. Third, BaBar has measured a
nonzero triple-product asymmetry in $B \to \phi K^*$ at $1.7\sigma$
\cite{BaBarTP}. However, this effect is expected to vanish in the SM
\cite{BVVTP}. Note that all of these new-physics signals should be
viewed with some skepticism. None of them is statistically
significant, and the two experiments have not yet converged on any of
the above measurements.  For example, while BaBar finds $\sin 2\beta
(\bd(t)\to \eta'\ks) = 0.27 \pm 0.14 \pm 0.03$ \cite{BaBaretap}, Belle
finds $\sin 2\beta (\bd(t)\to \eta'\ks) = 0.65 \pm 0.18 \pm 0.04$
\cite{Belleetap}, which is consistent with the value of $\sin 2\beta$
found in $\bd(t) \to J/\psi \ks$. Still, these hints are intriguing,
since all involve $B$ decays which receive large contributions from
$\btos$ penguin amplitudes. In this paper (as in Ref.~\cite{DLNP}), we
therefore adopt the point of view that there is new physics, and that
it contributes significantly only to those decays with sizeable
$\btos$ penguin amplitudes, but does not affect decays involving
$\btod$ penguins.

Consider now a $B\to f$ decay with a $\btos$ penguin. The NP
operators are assumed to be roughly the same size as the SM $\btos$
penguin operators, so that the new effects are important. There are
many potential NP operators. At the quark level, these take the form
${\cal O}_{\sss NP}^{ij,q} \sim {\bar s} \Gamma_i b \, {\bar q}
\Gamma_j q$ ($q = u,d,s,c$), where the $\Gamma_{i,j}$ represent
Lorentz structures, and colour indices are suppressed. The NP
contributes to $B\to f$ through the matrix elements $\bra{f} {\cal
O}_{\sss NP}^{ij,q} \ket{B}$. Each of the matrix elements can have
different weak phases and in principle each can also have a different
strong phase. However, it was argued in Ref.~\cite{DLNP} that all NP
strong phases are negligible, which leads to a great simplification:
one can combine all NP matrix elements into a single NP amplitude,
with a single weak phase:
\beq
\sum \bra{f} {\cal O}_{\sss NP}^{ij,q} \ket{B} = \ANPq e^{i \Phi_q} ~,
\label{NPparams}
\eeq
where $q=u,d,s,c$. (In general, the $\ANPq$ and $\Phi_q$ will be
process-dependent. The NP phase $\Phi_q$ will be the same for all
$\btos q {\bar q}$ decays only if all NP operators for the same
quark-level process have the same weak phase.)

While this argument --- that the new-physics strong phases are
negligible --- is quite general, there are still ways of getting
around this result. For example, it does not hold if the NP is quite
light, or if there is a significant enhancement of certain matrix
elements. The reader should therefore be aware of these caveats.

In Ref.~\cite{BNPmethods}, a number of methods for measuring the
$\ANPq$ and $\Phi_q$ were examined. Here we analyze another method. It
consists of using $\bd\to \pi^+\pi^-$ and $\bs\to K^+ K^-$
decays. These two decays, which are related by flavour SU(3), have
been used in the past to both obtain information about SM parameters
\cite{decaysSM,fl} and to detect the presence of new physics
\cite{decaysNP1,decaysNP2}. In all cases, one has to address the size
of SU(3)-breaking effects \cite{decaysSU3}. In the present paper, we
show that measurements of these two decays actually allow one to {\it
measure} the NP parameters $\ANPu$ and $\Phi_u$.

Consider the decay $\bs\to K^+ K^-$ within the SM. It is governed by
the quark-level process $\btos u {\bar u}$, and in terms of diagrams
\cite{GHLR}, the amplitude receives several contributions:
\beq
A(\bs \to K^+ K^-) = - T' - P' - E' - PA' - {2 \over 3} \pewcp ~.
\label{bsamp}
\eeq
In the above, the amplitude is written in terms of a colour-favored
tree amplitude $T'$, a gluonic penguin amplitude $P'$, an exchange
amplitude $E'$, a penguin annihilation amplitude $PA'$, and a
colour-suppressed electroweak penguin amplitude $\pewcp$. (The primes
on the amplitudes indicate a $\btos$ transition.) These various
contributions can be grouped into two types. There are charged-current
contributions, proportional to $V_{ub}^* V_{us}$, and the penguin-type
contributions $V_{ub}^* V_{us} P'_u + V_{cb}^* V_{cs} P'_c + V_{tb}^*
V_{ts} P'_t$. (Here the charged-current term includes $T'$ and $E'$,
while the penguin term includes $P'$, $PA'$ and $\pewcp$.) The
unitarity of the Cabibbo-Kobayashi-Maskawa (CKM) matrix can be used to
eliminate $V_{tb}^* V_{ts}$, so that the penguin-type contributions
become $V_{ub}^* V_{us} (P'_u - P'_t) + V_{cb}^* V_{cs} (P'_c -
P'_t)$. The amplitude for $\bs\to K^+ K^-$ can then be written as 
\cite{decaysSM}
\bea
A(\bs\to K^+ K^-) & = & V_{ub}^* V_{us} (A_{\sss CC}^{\prime u} +
A_{\rm pen}^{\prime ut}) + V_{cb}^* V_{cs} A_{\rm pen}^{\prime ct}
\nn\\
& = & \left\vert {V_{us} \over V_{ud}} \right\vert {\cal C}'
e^{i\theta_{\sss C'}} \left(e^{i \gamma} + \left\vert {V_{cs} V_{ud}
\over V_{us} V_{cd}} \right\vert d' e^{i\theta'} \right) ~.
\label{ampBsKK}
\eea
(Magnitudes of CKM matrix elements, such as $| V_{us}/V_{ud} |$, have
been included to permit easy comparison with the amplitude for the
decay $\bd\to\pi^+ \pi^-$ \cite{decaysSM,fl}). Here, $A_{\rm pen}^{\prime
 it} \equiv P'_i - P'_t$ ($i=u,c$), $\gamma$ is a SM CP-violating
phase \cite{pdg}, and \cite{decaysSM}
\bea
{\cal C}' e^{i\theta_{\sss C'}} & \equiv & |V_{ub}^* V_{ud}| \left( A_{\rm
CC}^{\prime u} + A_{\rm pen}^{\prime ut} \right) ~, \nn\\
d' e^{i\theta'} & \equiv & \frac{1}{R_b} \left( \frac{A_{\rm
 pen}^{\prime ct}} {A_{\sss CC}^{\prime u} + A_{\rm pen}^{\prime
 ut}} \right) ~,
\eea
with $R_b = |(V_{ub}^* V_{ud}) / (V_{cb}^* V_{cd}) |$. In the above,
$\theta'$ is the relative strong phase between $A_{\rm pen}^{\prime
ct}$ and $(A_{\sss CC}^{\prime u} + A_{\rm pen}^{\prime ut})$, and
$\theta_{\sss C'}$ is the strong phase associated with ${\cal C}'$. (If one
works entirely within the SM, $\theta_{\sss C'}$ is unimportant.)

Of course, since $\bs\to K^+ K^-$ involves a $\btos$ penguin
amplitude, one must include the effects of new physics. In this case,
only the $u$-quark NP parameters appear [see Eq.~(\ref{NPparams})].
Including these, the full amplitude for $\bs\to K^+ K^-$ can be
written
\beq
A(\bs\to K^+ K^-) = \left\vert {V_{us} \over V_{ud}} \right\vert {\cal
C}' e^{i\theta_{\sss C'}} \left(e^{i \gamma} + \left\vert {V_{cs} V_{ud}
\over V_{us} V_{cd}} \right\vert d' e^{i\theta'} \right) + \ANPu e^{i
\Phi_u} ~.
\label{BsKKNP}
\eeq
The amplitude for the CP-conjugate process, $\bsbar\to K^+ K^-$, can
be obtained from the above by simply changing the signs of the weak
phases $\gamma$ and $\Phi_u$.

There are a total of three measurements which can be made of
$\bs(t)\to K^+ K^-$: the total branching ratio, the direct CP
asymmetry, and the mixing-induced CP asymmetry. However, these
measurements depend on 8 theoretical parameters: ${\cal C}'$, $d'$,
$\theta_{\sss C'}$, $\theta'$, $\ANPu$, $\Phi_u$, $\gamma$, and the
$\bs$--$\bsbar$ mixing phase, $\phi_s$.

Two of these parameters can be measured independently. First, the weak
phase $\gamma$ can be obtained from $B$ decays which are unaffected by
new physics in $\btos$ penguin amplitudes. For instance, it can be
obtained from $B^\pm \to DK$ decays \cite{BDK}. Alternatively, the
angle $\alpha$ can be extracted from $B \to \pi\pi$ \cite{Bpipi},
$B\to\rho\pi$ \cite{Brhopi} or $B\to\rho\rho$ decays \cite{Brhorho}.
$\beta$ is already known from measurements of $\bd(t) \to J/\psi \ks$
\cite{ckmfit}, so that $\gamma$ can be obtained using $\gamma = \pi -
\beta - \gamma$.  Finally, it is also possible to take $\gamma$ from a
fit to the various unitarity-triangle measurements. (The present
preferred value is $\gamma = 62^\circ$ \cite{ckmfit,extra}.)

The second parameter is $\phi_s$, the phase of $\bs$--$\bsbar$
mixing. However, its measurement may pose theoretical problems. The
standard way to measure $\phi_s$ is through CP violation in $\bs(t)\to
J/\psi \eta$ (or $\bs(t)\to J/\psi \phi$). However, there is a
potential problem here: this decay receives NP contributions from
$O_{\sss NP}^c \sim \bar{s}b {\bar c} c$ operators (the Lorentz and
colour structures have been suppressed), so that there may be effects
from these NP operators in the measurement of $\phi_s$.

The solution to this problem can be found by considering
$\bd$--$\bdbar$ mixing. The phase of this mixing is unaffected by new
physics and thus takes its SM value, $\beta$. The canonical way to
measure this angle is via CP violation in $\bd(t)\to J/\psi \ks$.
However, this decay also receives NP contributions from $O_{\sss
NP}^c$ operators. On the other hand, the value of $\beta$ extracted
from $\bd(t)\to J/\psi \ks$ is in line with SM expectations.  This
strongly suggests that any $O_{\sss NP}^c$ contributions to this decay
are quite small, say $\lsim 20\%$.  Now, the non-strange part of the
$\eta$ wavefunction has a negligible contribution to $\bra{J/\psi
\eta}O_{\sss NP}^c \ket{\bs}$. Thus, this matrix element can be
related by flavour SU(3) to $\bra{J/\psi \ks}O_{\sss NP}^c \ket{\bd}$
(up to a mixing angle). That is, both matrix elements are very
small. In other words, we do not expect significant $O_{\sss NP}^c$
contributions to $\bs(t)\to J/\psi \eta$, so that $\phi_s$ can be
measured through CP violation in this decay, even in the presence of
NP. (Note that, in general, NP which affects $\btos$ transitions will
also contribute to $\bs$--$\bsbar$ mixing, i.e.\ one will have NP
operators of the form $\bar{s} b \, \bar{b} s$. In this case, the
phase of $\bs$--$\bsbar$ mixing may well differ from its SM value
($\simeq 0$) due to the presence of NP.)

Assuming that $\gamma$ and $\phi_s$ are measured independently, the
$\bs(t)\to K^+ K^-$ amplitude still depends on 6 unknown theoretical
quantities: ${\cal C}'$, $d'$, $\theta_{\sss C'}$, $\theta'$, $\ANPu$ and
$\Phi_u$. With only three experimental measurements, one cannot
extract the theoretical parameters. In particular, we cannot obtain
the NP quantities $\ANPu$ and $\Phi_u$.

The necessary additional information can be obtained by considering
measurements of $\bd \to\pi^+ \pi^-$. Within the SM, this decay is
related by flavour SU(3) to $\bs\to K^+ K^-$. However, because it is
described at the quark level by $\btod u {\bar u}$, it does not
receive NP contributions. Its amplitude can be written in terms of
diagrams as
\beq
A(\bd \to \pi^+ \pi^-) = - T - P - E - PA - {2 \over 3} \pewc ~,
\eeq
where the diagrams are written without primes to indicate a $\btod$
transition.

Similar to $\bs\to K^+ K^-$, the amplitude for $\bd \to\pi^+ \pi^-$
can be written in terms of pieces proportional to $V_{ub}^* V_{ud}$
and $V_{cb}^* V_{cd}$ \cite{decaysSM,fl,decaysNP1,decaysNP2}:
\bea
A(\bd\to \pi^+\pi^-) & = & V_{ub}^* V_{ud} (A_{\sss CC}^{u} + A_{\rm
pen}^{ut}) + V_{cb}^* V_{cd} A_{\rm pen}^{ct} \nn\\
& = & {\cal C} e^{i\theta_{\sss C}} \left(e^{i \gamma} - d e^{i\theta}
\right) ~,
\label{ampBpipi}
\eea
where $A_{\rm pen}^{it} \equiv P_i - P_t$ ($i=u,c$), and
\bea
{\cal C} e^{i\theta_{\sss C}} & \equiv & |V_{ub}^* V_{ud}| \left( A_{\rm
CC}^{u} + A_{\rm pen}^{ut} \right) ~, \nn\\
d e^{i\theta} & \equiv & \frac{1}{R_b} \left( \frac{A_{\rm pen}^{ct}}
{A_{\sss CC}^u + A_{\rm pen}^{ut}} \right) ~.
\eea
Since the strong phase $\theta_{\sss C}$ cannot be measured (it is just an
overall phase), the $\bd \to\pi^+ \pi^-$ amplitude depends on the four
quantities $\gamma$, ${\cal C}$, $d$ and $\theta$.

As with $\bs(t)\to K^+ K^-$, there are three measurements to be made
in $\bd(t)\to\pi^+\pi^-$: the total branching ratio, the direct CP
asymmetry, and the mixing-induced CP asymmetry. Thus, assuming that
$\gamma$ has been measured independently, the three
$\bd(t)\to\pi^+\pi^-$ measurements allow the extraction of ${\cal C}$,
$d$ and $\theta$. However, assuming a perfect flavour SU(3) symmetry,
we have ${\cal C}' = {\cal C}$, $d' = d$ and $\theta' = \theta$. With
this information, the number of unknown theoretical quantities in
$\bs(t)\to K^+ K^-$ is reduced to three: $\theta_{\sss C'}$ $\ANPu$ and
$\Phi_u$. The three measurements of $\bs(t)\to K^+ K^-$ therefore
permit us to obtain the parameters $\ANPu$ and $\Phi_u$.

This discussion shows that the measurements of $\bs(t)\to K^+ K^-$ and
$\bd(t)\to\pi^+\pi^-$, along with the independent determinations of
$\gamma$ and $\phi_s$, allow one to {\it measure} the NP quantities
$\ANPu$ and $\Phi_u$. The knowledge of these parameters, along with
those obtained via other methods \cite{BNPmethods} will allow us to
rule out many NP models and thus partially identify the new physics.

Of course, in practice, one has to examine the theoretical precision
with which $\ANPu$ and $\Phi_u$ can be obtained. In particular, one
has to take into account the SU(3)-breaking effects in relating ${\cal
C}'$, $d'$ and $\theta'$ to ${\cal C}$, $d$ and $\theta$. The ratio
$\vert {\cal C}'/{\cal C} \vert$ has recently been calculated using
QCD sum rules \cite{Mannel}:
\beq
\left\vert {{\cal C}' \over {\cal C}} \right\vert =
1.76^{+0.15}_{-0.17} ~.
\label{CprimeCratio}
\eeq
However, the effect of SU(3) breaking in relating the other parameters
has not yet been computed. In Ref.~\cite{decaysNP2}, the impact of
U-spin breaking was explored by allowing the parameters to vary in
certain ranges: $d'/d = 1.0 \pm 0.2$, and $\theta' - \theta = 0^\circ
\pm 40^\circ$. These uncertainties must be included in the extraction
of $\ANPu$ and $\Phi_u$. (Note: the assumed size of SU(3) breaking in
$d'/d$ and $\theta' - \theta$ is smaller than that found for $|{\cal
C}'/{\cal C}|$ in Eq.~(\ref{CprimeCratio}). This is because all
factorizable SU(3)-breaking effects cancel in the ratio $d'/d$,
leaving only nonfactorizable corrections. SU(3) breaking in $|{\cal
C}'/{\cal C}|$ is expected to be larger since both factorizable and
nonfactorizable terms are present.)

Suppose now that new physics has been found through measurements of
$\bd\to \pi^+\pi^-$ and $\bs\to K^+ K^-$, i.e.\ $\ANPu$ and $\Phi_u$
have been seen to be nonzero. Below we show how measurements of
$\bs\to K^0 \kbar$ can be used to partially identify the NP.

The SM amplitude for $\bs \to K^+ K^-$ is given in Eq.~(\ref{bsamp})
in terms of the diagrams $T'$, $P'$, $E'$, $PA'$ and $\pewcp$.
However, it is believed that $|E'|, |PA'|, |\pewcp| \ll |P'|$.
Neglecting these small pieces -- the theoretical error is expected to
be at the level of $\sim 5\%$ -- the decay amplitude becomes
\beq
A(\bs \to K^+ K^-) \simeq - T' - P' ~.
\eeq
This can be written as in Eq.~(\ref{BsKKNP})
(repeated here for convenience):
\beq
A(\bs\to K^+ K^-) = \left\vert {V_{us} \over V_{ud}} \right\vert {\cal
C}' e^{i\theta_{\sss C'}} \left(e^{i \gamma} + \left\vert {V_{cs} V_{ud}
\over V_{us} V_{cd}} \right\vert d' e^{i\theta'} \right) + \ANPu e^{i
\Phi_u} ~,
\eeq
where we have included the new-physics contribution. Here the SM
parameters (${\cal C}'$, $d'$, etc.) are related only to the diagrams
$T'$ and $P'$:
\bea
\left\vert {V_{us} \over V_{ud}} \right\vert {\cal C}'
e^{i\theta_{\sss C'}} e^{i \gamma} & \equiv & T' + V_{ub}^* V_{us} (P'_u -
P'_t) ~, \nn\\
\left\vert {V_{us} \over V_{ud}} \right\vert {\cal C}' d' e^{i (
 \theta_{\sss C'} + \theta')} & \equiv & V_{cb}^* V_{cs} (P'_c - P'_t)
 ~.
\eea
One can similarly neglect the small amplitudes in $\bd
\to \pi^+ \pi^-$, so that
\bea
A(\bd \to \pi^+ \pi^-) & \simeq & - T - P \nn\\
& = & {\cal C} e^{i\theta_{\sss C}} \left(e^{i \gamma} - d e^{i\theta}
\right) ~,
\eea
where again the SM parameters are related to $T$ and $P$ only. The
method described earlier still applies. Assuming that $\gamma$ and
$\phi_s$ are known independently, measurements of $\bs(t)\to K^+ K^-$
and $\bd(t)\to\pi^+\pi^-$ permit the extraction of all theoretical
quantities: the SM parameters, as well as $\ANPu$ and $\Phi_u$. This
is equivalent to measuring $T'$, $V_{ub}^* V_{us} (P'_u - P'_t)$ and
$V_{cb}^* V_{cs} (P'_c - P'_t)$, along with the NP parameters.

Now, the SM amplitude for $\bs \to K^0 {\bar K^0}$ is given by
\beq
A(\bs \to K^0 {\bar K^0}) = P' + PA' - {1 \over 3} \pewcp ~.
\eeq
As above, we can neglect $PA'$ and $\pewcp$. $P'$ also contains a
piece proportional to $V_{ub}^* V_{us}$. However, $|V_{ub}^* V_{us}
(P'_u - P'_t)| \ll |V_{cb}^* V_{cs} (P'_c - P'_t)|$ since $\left\vert
{V_{ub}^* V_{us} / V_{cb}^* V_{cs}} \right\vert \simeq 2\%$. We
therefore retain only the piece $V_{cb}^* V_{cs} (P'_c - P'_t)$. New
physics can affect this decay as well, but here only the $d$-quark NP
parameters appear:
\beq
A(\bs \to K^0 {\bar K^0}) = V_{cb}^* V_{cs} (P'_c - P'_t) + \ANPd e^{i
\Phi_d} ~.
\eeq
Note that $V_{cb}^* V_{cs} (P'_c - P'_t)$ is already known from the
$\bs(t) \to K^+ K^-$ analysis.

The measurement of $\bs(t) \to K^0 {\bar K^0}$ now allows us to
partially identify the new physics. If we assume that the NP is
isospin-conserving then $\ANPd = \ANPu$ and $\Phi_d = \Phi_u$. In
this case, the $\bs(t) \to K^+ K^-$ analysis gives us {\it all} of the
theoretical parameters in $A(\bs \to K^0 {\bar K^0})$, and we can
predict the values of the various measurable quantities in this decay
(the total branching ratio, the direct CP asymmetry, and the
mixing-induced CP asymmetry). We can then make measurements of $\bs(t)
\to K^0 {\bar K^0}$ and compare the results with the theoretical
predictions. If they disagree, we will have ruled out
isospin-conserving NP.

Particular types of new physics also allow us to establish the
amplitude for $\bs \to K^0 {\bar K^0}$. For example, suppose that
$Z$-mediated $\btos$ flavour-changing neutral currents are present. In
this case, we know how the $d$-quark NP parameters are related to the
$u$-quark NP parameters. In particular, $\Phi_d = \Phi_u$ and $\ANPd =
\ANPu \times c_d(Z)/c_u(Z)$, where $c_d(Z)$ and $c_u(Z)$ are the
(known) $Z$ couplings to ${\bar d} d$ and ${\bar u} u$, respectively.
Thus, for this model of NP, once again we know the full $\bs \to K^0
{\bar K^0}$ amplitude, and can compare experimental measurements with
theoretical predictions. A disagreement (beyond the level of the
theoretical uncertainty) will allow us to rule out this type of NP.

Finally, we examine the latest experimental results on
$\bd(t)\to\pi^+\pi^-$ and $\bs(t)\to K^+K^-$, and discuss their
implications for the method of measuring new-physics parameters
outlined earlier. We use data from BaBar \cite{BaBar}, Belle
\cite{Belle} and CDF \cite{punzi}.

We begin by presenting the SM expressions for $\av{{\rm BR}(\bs\to
K^+K^-)}$, $A_{dir}(\bs\to K^+K^-)$ and $A_{mix}(\bs\to K^+K^-)$ 
\cite{decaysSM}:
\bea 
\label{br}
& \av{{\rm BR}(\bs\to K^+K^-)}^{SM} = g_{ps} \, \epsilon \,
\mathcal{C}'^2 \Delta ~, & \\
\label{adir}
& A_{dir}^{SM}= \frac{\displaystyle 1}{\displaystyle\Delta} \left(
2\tilde{d}' \sin{\gamma} \sin{\theta'} \right) ~, & \\
\label{amix}
& A_{mix}^{SM}=\frac{\displaystyle 1}{\displaystyle \Delta} \left[
\sin{(2\gamma+\phi_s^{\sss SM})} + 2\tilde{d}' \cos{\theta'}
\sin{(\gamma+\phi_s^{\sss SM})}+ \tilde{d}'^2 \sin{\phi_s^{\sss SM}}
\right] ~. &
\eea
In the above, $\epsilon \equiv |V_{us}/V_{ud}|^2$, $\tilde{d}' \equiv
|(V_{cs} V_{ud}) / (V_{us} V_{cd})| d'$, $\Delta = 1 + 2 \tilde{d}'
\cos\gamma \cos{\theta'}+\tilde{d}'^2$ and $g_{ps} = \tau_{Bs}
\sqrt{1-4m_K^2/m_{\bs}^2}/(16 \, \pi \, m_{\bs})$.

In the presence of new physics, the deviations of these observables
from their SM expressions are given by
\bea
\label{eq:BKK}
& \av{{\rm BR}(\bs\to K^+K^-)} = \av{{\rm BR}(\bs\to
K^+K^-)}^{SM}\cdot (1+\mathcal{B}^{NP}) ~, & \\
& A_{dir}(\bs\to K^+K^-) = \frac{\displaystyle
A_{dir}^{SM}+\mathcal{D}^{NP}}{\displaystyle 1+\mathcal{B}^{NP}} ~, &
\\
& A_{mix}(\bs\to K^+K^-)= \frac{\displaystyle
A_{mix}^{SM}\cdot\cos{\delta\phi_s^{\sss
NP}}+\mathcal{M}^{NP}}{\displaystyle 1+\mathcal{B}^{NP}} ~, &
\eea
where $\phi_s=\phi_s^{SM}+\delta \phi_s^{NP}$ \cite{dem}, and
\bea
\mathcal{B}^{NP}& = & \frac{1}{\Delta}\big[ \textstyle z^2+
2z\big(\cos{\theta_{\sss C'}}\cos{(\Phi_u-\gamma)}+
\tilde{d}'\cos{\Phi_u}\cos{(\theta_{\sss C'}+\theta')}\big)\big] ~, \\
\mathcal{D}^{NP} & = & \frac{1}{\Delta}\big[ 2z
(\tilde{d}'\sin{\Phi_u}\sin{(\theta_{\sss C'}+\theta')}+
\sin{\theta_{\sss C'}}\sin{(\Phi_{u}-\gamma)})\big] ~, \\
\mathcal{M}^{NP}&=&\frac{1}{\Delta}\big[ z^2\sin{(2\Phi_u+\phi_s)}+
 2z(\cos{\theta_{\sss C'}}\sin{(\Phi_u+\phi_s+\gamma)} \cr \nn \\
&+& \tilde{d}'\cos{(\theta_{\sss C'}+\theta')}\sin{(\Phi_u+\phi_s)})
 \\
&+&{\left( \cos{(2\gamma+\phi_s^{\sss SM})}+
2\tilde{d}'\cos{\theta'}\cos{(\gamma+\phi_s^{\sss SM})}+
\tilde{d}'^2\cos{\phi_s^{\sss SM}} \right) \sin{\delta\phi_s^{\sss
NP}}}\big] ~, \nn
\eea
with $z={\cal A}^u/(\sqrt{\epsilon}~{\cal C}^\prime)$. 
In order to measure these NP functions, one needs predictions for
$A_{dir}^{SM}(\bs\to K^+K^-)$, $A_{mix}^{SM}(\bs\to K^+K^-)$ and ${\rm
BR}^{SM}(\bs\to K^+K^-)$. As described earlier, this information can
be obtained from measurements of $\bd(t) \to \pi^+\pi^-$. To be
precise, these measurements allow one to extract $d$, $\theta$ and
${\cal C}$, which then allow us to obtain the SM predictions for the
$\bs\to K^+ K^-$ observables, given a prescription for the inclusion
of U-spin breaking effects.

The present experimental situation of the mixing induced and direct CP
asymmetry of $\bd \to \pi^+\pi^-$ is still quite uncertain
\cite{BaBar,Belle}:
\beq
{A}_{dir}(\bd\to\pi^+\pi^-)= \left\{
\begin{array}{ll}
-0.09\pm0.15\pm0.04 & \mbox{(BaBar)}\\
-0.58\pm0.15\pm0.07 & \mbox{(Belle )}\nn
\end{array}
\right.
\eeq
\beq
{A}_{mix}(\bd\to\pi^+\pi^-)=\left\{
\begin{array}{ll}
0.30\pm0.17\pm0.03& \mbox{(BaBar)}\\
1.00\pm0.21\pm0.07 & \mbox{(Belle)} \nn
\end{array}
\right.
\eeq
\beq \label{databr} \langle{\rm BR}(\bd \to \pi^+\pi^-)\rangle= \left\{
\begin{array}{ll}
(4.7\pm0.6\pm0.2) \times 10^{-6} & \mbox{(BaBar)}\\
(4.4\pm0.6\pm0.3) \times 10^{-6} & \mbox{(Belle)} \nn
\end{array}
\right.
\eeq
In light of this, we perform the analysis separately for the BaBar and
Belle data.

The range of values for $d$ and $\theta$ are obtained as in \cite{fl}
and corresponds to the shaded regions in Fig.~1a (BaBar) and Fig.~1b
(Belle). The CKM angle $\gamma$ is taken to be in the range $\gamma
(degrees)=62^{+10}_{-12}$ \cite{ckmfit} and the $\bd$--$\bdbar$ weak
mixing phase is set to $\phi_d=48^\circ$ \cite{ckmfit}.

{}From the BaBar experimental values (within $\pm 1\sigma$) we find
two separate sets of solutions for $d$ and $\theta$ which are
compatible with ${A}_{mix} (\bd\to\pi^+\pi^-)$ and ${A}_{dir} 
(\bd\to\pi^+\pi^-)$. The first solution gives an
unconstrained range for $\theta$, while the allowed values for $d$
are:
\beq
 d\in (0,0.47) \label{dbabar}
\eeq
The value of the parameter ${\cal C}$ can be obtained from
Eqs.~(\ref{br}), (\ref{databr}) and (\ref{dbabar}), and lies in the
following range:
\beq
\mathcal{C}\in (1.7,2.8) ~\times 
10^{-8}~{\rm GeV}~.
\label{Cbabar}
\eeq
The second solution has $d>2.5$, which is strongly disfavoured by
QCD-factorization arguments\cite{ser} which predicts a value around
$0.3$ \cite{bn}. For this reason, we do not consider this second
solution further. Moreover, in \cite{fl} it was shown that these large
values of $d$ give rise to complex solutions when the ratio of neutral
channels are taken into account and are therefore unphysical.

Turning to the Belle data, we obtain a very large range for $d$, which
also includes quite unrealistic values. However, since these values
are not split into two regions, here we will take the complete range.
The allowed values for the hadronic parameters with the Belle data
are:
\beq
\begin{array}{l}
d\in(0.51,3.95) ~, \\ 
\theta\in(102^\circ,147^\circ) ~,
\label{dbelle} 
\end{array}
\eeq
and the corresponding range for $\mathcal{C}$ is given by:
\beq
\mathcal{C}\in (0.5,1.9)\times 10^{-8} ~{\rm GeV}~.
\label{Cbelle}
\eeq

The predicted values for the mixing-induced and direct CP asymmetries
for $\bs(t)\to K^+ K^-$ in the SM are given in Table 1 on the basis of
the BaBar and Belle data. These ranges take into account U-spin
breaking effects, as described in Eq.~(\ref{CprimeCratio}) and the
discussion following it. They also include the correlations between
$d$ and $\theta$ for each value of $\gamma$ implied by the shaded
regions in Figs.~1a and 1b. While the prediction from present Babar
data is compatible with all possible values for the $\bs(t)\to K^+
K^-$ asymmetries, Belle gives quite constrained ranges. In
Ref.~\cite{fl} the average value of these asymmetries with previous
data was also computed.

\vspace{-0.5cm}
\begin{figure} 
\label{fig}
\begin{center}
 \begin{minipage}[t]{6cm}
 \psfrag{d}[rb][rb][1][-90]{$d\ $}
 \psfrag{t}[c]{$\theta\;(rad)$}
 \psfrag{a}{$a)$}
 \psfrag{b}{$b)$}
 \psfrag{0}[]{$\scriptstyle 0$} \psfrag{1}[]{$\scriptstyle 1$}
\psfrag{2}[]{$\scriptstyle 2$}
 \psfrag{3}[]{$\scriptstyle 3$}\psfrag{4}[]{$\scriptstyle 4$}
\psfrag{5}[]{$\scriptstyle 5$}
 \psfrag{6}[]{$\scriptstyle 6$}\psfrag{-1}[]{$\scriptstyle -1$}
 \vspace{-4pt}
 \centering
 \includegraphics[width=6cm,height=6.5cm]{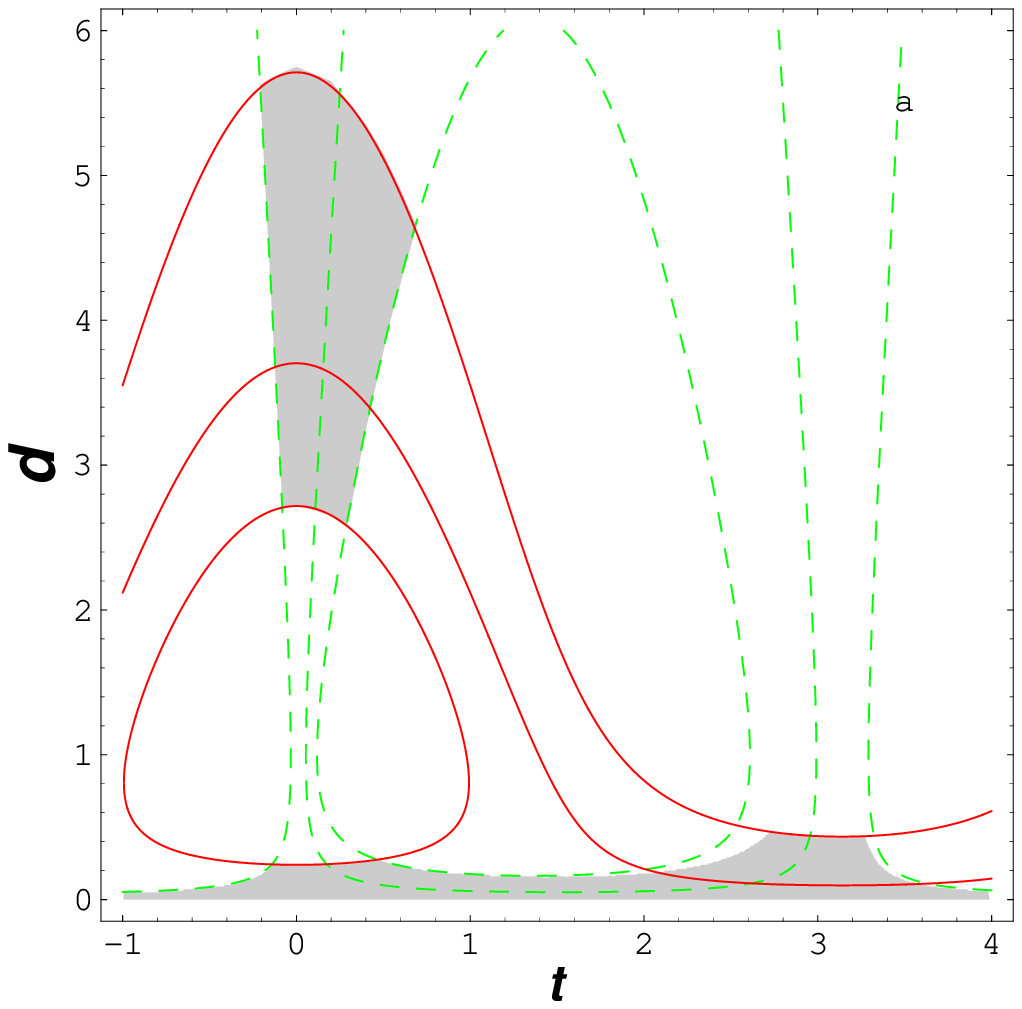}
 \end{minipage}%
 \begin{minipage}[t]{7cm}
 \psfrag{d}[][rB][1][-90]{$d$}
 \psfrag{t}[cb]{$\theta\;(rad)$}
 \psfrag{a}{$a)$}
 \psfrag{b}{$b)$}
 \psfrag{0}{$\scriptstyle 0$} \psfrag{1}{$\scriptstyle 1$}
\psfrag{2}{$\scriptstyle 2$}
 \psfrag{3}{$\scriptstyle 3$}\psfrag{4}{$\scriptstyle 4$}
\psfrag{5}{$\scriptstyle 5$}
 \psfrag{6}{$\scriptstyle 6$}\psfrag{-1}{$\scriptstyle -1$}
 \vspace{-6pt}
 \centering
 \includegraphics[width=6.5cm,height=6.75cm]{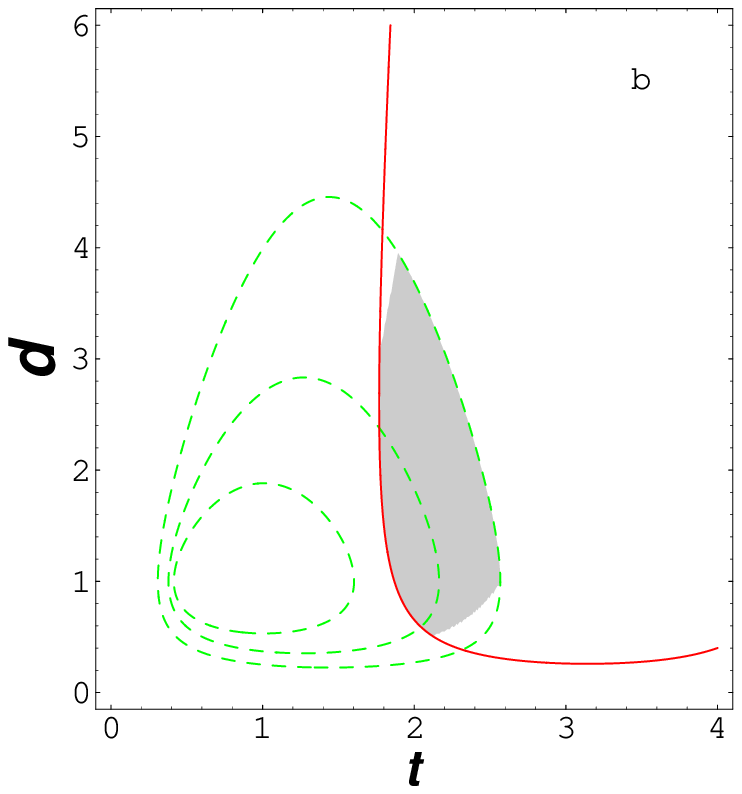}
 \end{minipage}
 %\begin{minipage}[t]{12cm}
 %\noindent \footnotesize \textit
\caption{a) The dashed lines show the allowed region in the $d-\theta$
plane using only BaBar experimental data on
$A_{dir}(\bd\to\pi^+\pi^-)$. We allow for $\pm 1\sigma$ deviations of
$A_{dir}(\bd\to\pi^+\pi^-)$ and $\gamma$. The solid lines show the
same for the measured value of $A_{mix}(\bd\to\pi^+\pi^-)$, also at
$\pm 1\sigma$. The shaded regions are those values consistent with
both observables for each value of $\gamma$. b) Same for Belle 
experimental data.}
%\end{minipage}
\end{center}
\end{figure}

\vskip5truemm

At present, the observables ${A}_{dir}(\bs\to K^+K^-)$, ${A}_{mix}(\bs\to 
K^+K^-)$ have not yet been measured, so we cannot give
estimates for ${\cal D}^{NP}$ and ${\cal M}^{NP}$. As for ${\rm
BR}(\bs\to K^+K^-)$, it is more useful to present the ratio of
branching ratios of $\bs\to K^+K^-$ and $\bd\to \pi^+\pi^-$: $R_{d}^{s
\; SM} \equiv BR(\bs \to K^+ K^-)/BR(\bd \to \pi^+ \pi^-)$
\cite{decaysNP2}. This ratio, which has been measured experimentally
\cite{punzi}, does not require in the SM the knowledge of ${\cal C}$
and ${\cal C}^\prime$ individually, but only the ratio, which can be
taken from Eq.~(\ref{CprimeCratio}). Our prediction for this quantity
in the SM is:
\bea
& 0.1 \leq R_{d \; BaBar}^{s \; SM} \leq 13.9 ~, & \nn\\
& 3.6 \leq R_{d \; Belle}^{s \; SM} \leq 85.5 ~. &
\eea
This can be compared with a recent experimental measurement
at CDF \cite{punzi}:
\beq
\frac{f_d}{f_s}\frac{BR(\bd\to \pi^+\pi^-)}
{BR(\bs\to K^+K^-)}=0.48\pm 0.14 ~,
\label{BRpBRs}
\eeq
which implies
\beq
R_d^s = 7.7\pm 2.5 ~.
\eeq

We therefore see that, with the present data, the SM prediction for
$R_d^s$ overlaps with the experimental range for both experiments
BaBar and Belle. There is therefore no evidence for a nonzero
$\mathcal{B}^{NP}$. However, the very large errors leave much room for
new physics. These errors will be dramatically reduced with improved
experimental measurements for the $\bd\to \pi^+\pi^-$ observables.

\begin{table}[t] \label{tab}
\begin{center}
\begin{tabular}{|c|c|c|}
\hline
Observable & BaBar & Belle \\
\hline ${A}_{dir}^{SM}(\bs\to K^+K^-)$ & (~$-$1~,~1~) & 
(~$-$0.02~,~0.18~) \\
${A}_{mix}^{SM}(\bs\to K^+K^-)$ & (~$-$1~,~1~) & (~$-$0.19~,~0.05~) \\
\hline
\end{tabular}
\caption{Predicted ranges for the $\bs(t)\to K^+ K^-$ CP asymmetries
in the SM using Eqs.~(\ref{br})-(\ref{amix}). U-spin breaking effects
have been included as explained in the text.}
\label{tab:BPIK-obs}
\end{center}
%\vspace*{-0.3truecm}
\end{table}

%We can obtain from together with the theoretical prediction for
%the $R_s^{d \; SM}$, using Eq.~(\ref{yy}):
%$$
%{\cal B}^{NP}=\frac{R_s^{d \;exp}}{R_s^{d \;SM}}-1
%$$ and including all U-spin breaking effects:

%begin{equation}
%\left\{ \begin{array}{l}
%   \mathcal{B}^{NP}_{BaBar}\in (aa,bb)\\
%   \mathcal{B}^{NP}_{Belle}\in (-0.97,5.39)
%  \end{array} \right.
%\eeq

To summarize, present data hints at new physics in $\btos$ penguin
amplitudes, but there is no evidence for NP in decays involving
$\btod$ penguins. We therefore assume that the NP is present only in
$\btos$ transitions. Recently it was shown that the NP strong phases
are negligible \cite{DLNP}, in which case all NP effects can be
parametrized in terms of a single amplitude $\ANPq$ and phase $\Phi_q$
for each quark-level decay $\btos q {\bar q}$ ($q=u,d,s,c$). In this
paper we have shown that one can {\it measure} the NP parameters
$\ANPu$ and $\Phi_u$ using the decays $\bs(t)\to K^+ K^-$ and
$\bd(t)\to \pi^+\pi^-$. Because these processes are related by flavour
SU(3), there is some theoretical uncertainty in the method. We have
shown explicitly how this technique can be implemented experimentally.

We have also shown that, given the presence of nonzero $\ANPu$ and
$\Phi_u$ in $\bs(t)\to K^+ K^-$, meaurements of the decay $\bs(t)\to
K^0 \kbar$ can be used to partially identify the NP. This latter decay
involves the NP parameters $\ANPd$ and $\Phi_d$. Within certain
classes of NP models, there is a known relationship between $\ANPu$
and $\Phi_u$ and $\ANPd$ and $\Phi_d$. This allows us to {\it predict}
the values of quantities in $\bs(t)\to K^0 \kbar$, given measurements
of $\bs(t)\to K^+ K^-$ and $\bd(t)\to \pi^+\pi^-$. Should the
experimental values of these quantities disagree with the predictions,
this would rule out these NP models.

Finally, we have discussed the implications of present data for the
method of measuring $\ANPu$ and $\Phi_u$ in $\bs(t)\to K^+ K^-$ and
$\bd(t)\to \pi^+\pi^-$. Although we have measurements of the various
$\bd(t)\to \pi^+\pi^-$ observables, albeit with large (and sometimes
inconsistent) errors, only the branching ratio for $\bs\to K^+ K^-$
has been measured. The comparison of the branching ratios for these
two processes does not suggest the presence of NP. However, the errors
are still very large, leaving much space for this NP. In the future,
these errors will be greatly reduced with improved experimental
measurements.

\bigskip
\noindent
{\bf Acknowledgements}: \\
%\bigskip
We thank G. Punzi for helpful discussions.
DL thanks JM for the hospitality of the Universitat Aut\`onoma de
Barcelona, where part of this work was done. This work was financially
supported by NSERC of Canada (DL) and by FPA2002-00748 (JM \& JV) and the 
Ramon y
Cajal Program (JM).

%%%%%%%%%%%%%%%%%%%%% REFERENCES %%%%%%%%%%%%%%%%%%%%%%%%%%%%%%%%

\end{document}